# What Way Is It Meant To Be Played?

Florian Mihola

March 2020

## Abstract

The most commonly used interface between a video game and the human user is a handheld "game controller", "game pad", or in some occasions an "arcade stick." Directional pads, analog sticks and buttons—both digital and analog—are linked to in-game actions. One or multiple simultaneous inputs may be necessary to communicate the intentions of the user. Activating controls may be more or less convenient depending on their position and size. In order to enable the user to perform all inputs which are necessary during gameplay, it is thus imperative to find a mapping between in-game actions and buttons, analog sticks, and so on. We present simple formats for such mappings as well as for the constraints on possible inputs which are either determined by a physical game controller or required to be met for a game software, along with methods to transform said constraints via a button-action mapping and to check one constraint set against another, i.e., to check whether a button-action mapping allows a controller to be used in conjunction with a game software, while preserving all desired properties.

## 1 Introduction

From the earliest days of arcade games up till present day, a variety of input devices have been explored. Pong uses paddles, Pac-Man moves in four directions via a four-way stick; there are steering wheels and pedals, light guns, and even skis to ride on. But apart from so-called "dedicated cabinets"—which use controls which are tailored to a specific game—most arcade cabinets use a digital eight or four-way stick—most often alongside a number of digital buttons. In home computers and home video game consoles digital inputs were the standard up until the "16-bit" era of the 1990s. Sony PlayStation, Nintendo 64 and Sega Saturn are among the first which brought with them *additional* analog controls—either at launch or as an updated controller option. And even though modern mass-market offerings include analog sticks and analog triggers, digital buttons and directional pads remain the ubiquitous fundamentals of input. The simple nature and widespread use of digital inputs leads to a degree of interoperability: Game software is not necessarily tied to a single game controller—whether we interpret this as a specific model, a design and protocol available by different manufacturers, or a class of generic controllers—but can be enjoyed using a range of controllers, provided they share at least some common characteristics. When a game releases for multiple contemporary platforms, differences in controls tend to be limited to different naming conventions; emulation of a so-called "retro game" most often calls for a minimum number of inputs; on platforms for which no single standard controller exists an appropriate controller choice and configuration are frequently left to the user.

But the decision which button mapping to use or whether a pair of controller and game "fit" at all is not as straight-forward as it might seem—especially so if we aim for more than the mere feasibility of gameplay.

How should controllers, games, and button mappings be encoded and how do we decide whether an instance of all three combined will deliver an enjoyable experience?

The rest of the paper explores how it relates to other work in the general area of interest in section 2, gives an introduction to our method—section 3—followed by a more formal explanation—section 4—and some insights into software implementations—section 5. Sections 6 and 7 discuss a variation on, and an expansion to our method and sections 8 looks back as well as forward—concluding what has been learned and anticipating future research.

## 2 Related Work

Reasonings along the lines laid out here, or methods similar to those presented, are likely to have been employed in practice, and one may very well deliver excellent playability of video games either with a fixed control scheme, or in the case of games which do not put stringent requirements on button configuration, or simply do not use a great number of inputs. We nevertheless maintain that, in order to properly discuss such techniques, spread their use in game and emulation software, and to build a common vocabulary, it is necessary to give accurate accounts of models and related algorithms in an academic setting.

Most of the sources we discovered barely overlap with our specific topic, but there is a range of works which are of interest to the same audience:

Skalski et al. [10] examine how *natural mappings*, using controllers like Nintendo's Wii motion controller affect game enjoyment. While the intersection of controllers allowing for such natural mappings and the ones in the focus of our method is small, we pursue similar goals of furthering game enjoyment and ease of playability. Another investigation into natural mappings by Shafer, Carbonara, and Popova even poses the question "Controller Required?" [9]; this is clearly related to—but also very different from—our interest in traditional gamepads. Seemingly in contrast, a study by Rogers, Bowman, and Oliver [8] alludes to situations where a traditional control scheme may in some aspects outperform motion controls.

Alves et al. [2] apply so-called "Design Thinking" techniques to designing novel game controllers; some designers consider hybrid approaches where physical inputs are combined with touch screen controls.

Button layouts—not of physical controllers, but virtual layouts for games played on touch screen devices—and their generation are discussed by Alves, Montenegro, and Trevisan [1] and Baldauf et al. [3]. The later article mentions Nintendo's Game & Watch series of handhelds—which use the kinds of digital inputs we are most interested in; reminding us of an interesting case study, albeit with rather simple controls.

Go, Konishi, and Matsuura [4] describe "IToNe", an application of dual analog joysticks for entering japanese text. It might be feasible to model this approach of partitioning a joysticks range into zones corresponding to japanese *kana* using our approach; it remains to be seen what, if anything, of interest could be added in this vein.

## 3 Overview

Guided by motivating examples, we will first give a somewhat informal introduction to the central ideas and methods of this paper, followed by formal definitions in section 4.

Starting with a simple notion, consider the number of inputs required to play a game in relation to the number of inputs found on an input device. In a 2D platformer we might find the game to react to four directional inputs and an additional jump command. A four-way directional pad and a button should suffice. The games requirements and a simple controller are succinctly represented as sets $\mathbb{G}$ and $\mathbb{C}$.

$$\mathbb{G} = \{left, right, up, down, jump\}$$
$$\mathbb{C} = \{dpad_{left}, dpad_{right}, dpad_{up}, dpad_{down}, button_a, button_b\}$$

Via a mapping of buttons to game controls such as $\{dpad_{left} \to left, dpad_{right} \to right, dpad_{up} \to up, dpad_{down} \to down, button_a \to jump\}$. It can be swiftly argued that it is possible to perform all desired in-game actions described by $\mathbb{G}$ using the controller described by $\mathbb{C}$ because if we replace every element of $\mathbb{G}$ with the corresponding element in $\mathbb{C}$ we obtain exactly set $\mathbb{G}$, meaning all game controls are connected to controller inputs in some way. Expanding on this example, assume the game character has the additional ability to run at different speeds, which shall be controlled by holding down a $run$ button.[1] Instead of simply expanding the previously mentioned sets and mapping we opt for a more sophisticated approach. In order to perform longer or higher jumps it is necessary for the player to not only have access to a $jump$ and $run$ button, it is paramount that the user is able to push the two corresponding buttons *at the same time*. In addition we want to express that it should *not* be possible to activate simultaneously both $left$ and $right$ or both $up$ and $down$. Along with the constraints imposed on the game's controls, the structure of the sets representing both game requirements and controller facilities increases in complexity. Sets like $\{jump, run\}$ or $\{right, run\}$ form the *elements* of what is now a set of sets representing all combinations of inputs which shall be possible to be activated *simultaneously*. We arrive at set

$$\begin{aligned}
\mathbb{G}_2 = \{&\{jump\}, \{run\}, \{jump, run\}, \\
&\{left\}, \{left, jump\}, \{left, run\}, \\
&\{left, jump, run\}, \{left, up\}, \\
&\{left, up, jump\}, \{left, up, run\}, \\
&\{left, up, jump, run\}, \{left, down\}, \\
&\{left, down, jump\}, \{left, down, run\}, \\
&\{left, down, jump, run\}, \\
&\{right\}, \{right, jump\}, \{right, run\}, \\
&\{right, jump, run\}, \{right, up\}, \\
&\{right, up, jump\}, \{right, up, run\}, \\
&\{right, up, jump, run\}, \{right, down\}, \\
&\{right, down, jump\}, \{right, down, run\}, \\
&\{right, down, jump, run\}, \\
&\{up\}, \{up, jump\}, \{up, run\}, \\
&\{up, jump, run\}, \\
&\{down\}, \{down, jump\}, \{down, run\}, \\
&\{down, jump, run\}\}
\end{aligned}$$

---

[1] It is not by accident that this control schemes resembles the one found in Super Mario Bros.

Set $\mathbb{C}_2$ is just as verbose, but offers no new insights and is thus omitted.[2] As for the button mapping, the changes are not as dramatic; we only need to add a suitable mapping for the $run$ action: $\{dpad_{left} \to left, dpad_{right} \to right, dpad_{up} \to up, dpad_{down} \to down, button_a \to jump, button_b \to run\}$. Via substitution according to this mapping we again arrive at exactly the set of game requirements, $\mathbb{G}_2$. Instead of stating that we need to be able to use various input combinations containing $left$ or $right$—but never both—we could also opt for the converse approach and specify that $left$ and $right$—likewise $up$ and $down$—build sets of buttons (or of in-game actions respectively) which may not be pushed (active) simultaneously. We may further encode that a button shall be operated with the left—or right—hand, that two actions shall be mapped to a pair of buttons found in a symmetrical configuration on a two-hand controller, or the organization into "face buttons" versus "trigger buttons". One may need to employ multiple such *predicates* to adequately represent all desired qualities, and it shall sometimes be beneficial to replace the "naked" elements employed so far with a predicate notation like $Facebutton(jump)$ or $Simultaneous(button_a, button_b)$. Note that we do allow both mapping multiple buttons to the same in-game action, as well as mapping a single button to multiple actions which shall be triggered simultaneously. A button may also not be mapped to anything at all. As long as all requirements stated as a game's predicates are included in those predicates corresponding to a controller—after all of them have been converted using a *single* mapping—the game and controller shall be considered compatible.

## 4 Definitions

This section shall describe in a formal way those structures which are essential for the proposed method.

**Definition 1.** Let $\mathbb{B}$ be the set of boolean values $\{true, false\}$ and let $\mathcal{P}(\mathbb{A})$ denote the powerset of set $\mathbb{A}$.

Then, if set $\mathbb{C}$ is a set of scalar values and $P : \mathcal{P}(\mathbb{C}) \mapsto \mathbb{B}$ is a function from subsets of $\mathbb{C}$ to either $true$ or $false$ we shall call $\mathbb{C}$ the *set of constants* associated with the *predicate* $P$. We denote by $\mathcal{S}(P)$ the *predicate set* of $P$, i.e., the subset of $\mathcal{P}(\mathbb{C})$ such that for each $x \in \mathcal{P}(\mathbb{C})$, $P(x) = true$ if and only if $x \in \mathcal{S}(P)$. It follows that if $P(x) = false$ then $x \notin \mathcal{S}(P)$. Note that $\mathcal{P}(\mathbb{C})$ includes $\varnothing$, the empty set.

In predicate logic, the elements of $\mathbb{C}$ would be referred to as *constants* or *nullary functions*. We will not consider any "proper" functions which take one or more arguments.

---

[2] More space-efficient ways to describe sets like these are introduced in section 6.

As we have seen previously, explicit lists of a predicate set can become rather unwieldy.

**Definition 2.** Let $simultaneously : \mathcal{P}(\mathbb{A}) \times \mathcal{P}(\mathbb{A}) \mapsto \mathcal{P}(\mathbb{A})$ be the function, polymorphic over $\mathbb{A}$, which maps two sets (of sets) $\mathbb{R}$ and $\mathbb{S}$ to the set of all sets which are the union of an element of $\mathbb{R}$ and an element of $\mathbb{S}$. That is to say, $x \in simultaneously(R, S)$ if and only if there are elements $r \in \mathbb{R}$ and $s \in \mathbb{S}$ such that $x = r \cup s$.

Using $simultaneously$, predicate sets representing inputs which are independent from each other may be combined without listing all combinations explicitly: Assuming just three actions $left$, $right$ and $shoot$, the combination of movement—$\{\varnothing, \{left\}, \{right\}\}$—and actions—$\{\varnothing, \{shoot\}\}$—may be stated as $simultaneously(\{\varnothing, \{left\}, \{right\}\}, \{\varnothing, \{jump\}\})$ which is equal to the more verbose $\{\varnothing, \{left\}, \{right\}, \{shoot\}, \{left, shoot\}, \{right, shoot\}\}$. When more, or larger, predicates are combined the effect can become much more pronounced.

In an expressive language like Haskell we can combine $simultaneously$ with a few constants and simple functions—which we call *combinators*—which may be seen as a simple embedded language, as shown in Fig. 1.

Mappings are sets of ordered pairs, but a different syntax gives a better hint as to their meaning.

**Definition 3.** Let $\mathbb{R}$ and $\mathbb{S}$ be two sets with $\mathbb{R} \cap \mathbb{S} = \varnothing$. We denote ordered pairs by $r \to s$ and a mapping from $\mathbb{R}$ to $\mathbb{S}$ by $\mathcal{M}(\mathbb{R}, \mathbb{S}) \subset \{r \to s | r \in \mathbb{R}, s \in \mathbb{S}\}$.

A mapping $\mathcal{M}(\mathbb{R}, \mathbb{S})$ may contain pairs $r_1 \to s_1$ and $r_2 \to s_2$ where either $r_1 = r_2$ or $s_1 = s_2$ in order to map a single button to multiple game inputs or to map multiple buttons to the same input.

Now that we know how to represent predicates and mappings it remains to show methods for first mapping a predicate from one set of constants—for example, buttons on a controller—to another—which may be controls in a game, or the buttons of another controller—and then to check whether one predicate set contains another.

```haskell
simultaneously xs ys = Set.fromList
  (Set.union <$> (Set.toList xs)
             <*> (Set.toList ys))
noInput = Set.singleton Set.empty
or      = Set.union
choose1 = Set.map Set.singleton
        . Set.fromList
atMost1 = (or noInput) . choose1
maybe x = noInput or Set.singleton
                    (Set.singleton x)

data Actions = Left  | Right | Up
             | Down  | Jump  | Run

jumpman = atMost1 [Left, Right]
  simultaneously atMost1 [Up, Down]
  simultaneously maybe Jump
  simultaneously maybe Run
```

Figure 1: Haskell definitions and predicate "jumpman" which is equivalent to $\mathbb{G}_2$ shown previously. Type signatures are omitted.

**Definition 4.** We denote by $map$ a function which maps one predicate set to another via a mapping. Let $\mathcal{M}(\mathbb{R}, \mathbb{S})$ be a mapping and $P : \mathcal{P}(\mathbb{R}) \mapsto \mathbb{B}$ be a predicate.

Then $\mathcal{S}(P) = \{x | x \in \mathcal{P}(\mathbb{R}) \land P(x) = true\}$ and $P' : \mathcal{P}(\mathbb{S}) \mapsto \mathbb{B}$ is a predicate with

$$\mathcal{S}(P') = map(\mathcal{S}(P), \mathcal{M}(\mathbb{R}, \mathbb{S}))$$
$$= \left\{ \bigcup_{r \in \mathbb{R}'} \{s | \exists r \to s \in \mathcal{M}(\mathbb{R}, \mathbb{S})\} \middle| \mathbb{R}' \in \mathcal{S}(P) \right\}$$

When multiple buttons are mapped to the same in-game action, multiple elements in $\mathcal{S}(P)$ may result in the same element in $\mathcal{S}(P')$ and when one button maps to multiple in-game actions we may need to be careful so as not to make it impossible to construct sets which contain *some but not all* of those in-game actions.

Finally, we need to establish what is necessary for a triplet containing two predicates and one mapping to be considered, as we say, *valid*, e.g., to enable one to use a controller and a mapping to play a game the way it is meant to be played.

**Definition 5.** Let $P$ and $P'$ be two predicates over a set of constants $\mathbb{C}$.

Then, if $\mathcal{S}(P) \subset \mathcal{S}(P')$, that is, if the predicate set of $P$ is a subset of the predicate set of $P'$ we say that $P$ is a *sub-predicate* of $P'$.

As this only allows us examine the relationship of two predicates over the *same* set of constants we need to expand the previous definition to predicates of *non-overlapping* constant sets:

**Definition 6.** Let $P$ and $P'$ be two predicates over two sets of constants $\mathbb{C}$ and $\mathbb{C}'$, respectively, where not only $P \neq P'$ but also $\mathbb{C} \cap \mathbb{C}' = \varnothing$ and let $M = \mathcal{M}(\mathbb{C}', \mathbb{C})$ be a mapping from $P'$ to $P$.

Then, if $\mathcal{S}(P) \subset map(\mathcal{S}(P'), M)$, that is, if the predicate set $P$ is a subset of the predicate set $P'$ mapped via $M$ we say that $P$ is a *sub-predicate* of $P'$ modulo $M$.

## 5  Implementation

Before expanding on it further, we will discuss one property that was crucial in the design of the approach discussed so far: relative ease of implementation. As will be shown in section 6, representing a controller's constraints—or game's requirements—as a set of sets does not constitute the most efficient use of memory space; there is in a fact a very straight-forward way to "compress" it, or, if viewed differently, there is an obvious redundancy which is covered in section 6.

From a practical standpoint, and with a desire to promote the inclusion of an algorithm or practice, how do we entice developers to adopt some form of it in future creations? A piece of free, open source source code is an obvious choice, but, alas, there are some drawbacks. For example, which language should it be written in? Any one choice would be limiting the number of platforms it could be applied to; or would have unacceptable performance in a low-level environment; or would be unable to handle the specific problem at hand; or would have multiple of these as well as other problems.

We choose thus, to explain briefly how to simply—naively, in a positive sense—implement the non-optimized method of using sets of sets. All that is needed are a few simple operations such as *bitwise and* and *or*, comparison of machine words, and some of the most fundamental control flow

```c
// assume 16 bit words
typedef uint16_t wt;

#define GAME_N 5
enum Game { GLeft = 1<<0, GRight = 1<<1,
  GUp = 1<<2, GDown = 1<<3, GJump = 1<<4
};

#define CONTROLLER_N 6
enum Controller { CLeft = 1<<0,
  CRight = 1<<1, CUp = 1<<2,
  CDown = 1<<3, CA = 1<<4, CB = 1<<5
};

#define PREDICATE_N 29
wt predicate[PREDICATE_N] = {
  CLeft,
  CLeft | CUp,
  CLeft | CUp | CA,
  CLeft | CUp | CB,
  CLeft | CUp | CA | CB,
  CLeft | CDown,
  CLeft | CDown | CA,
  CLeft | CDown | CB,
  CLeft | CDown | CA | CB,
  CRight,
  CRight | CUp,
  CRight | CUp | CA,
  CRight | CUp | CB,
  CRight | CUp | CA | CB,
  CRight | CDown,
  CRight | CDown | CA,
  CRight | CDown | CB,
  CRight | CDown | CA | CB,
  // some lines omitted
};

#define MAPPING_N 6
wt mapping[MAPPING_N * 2] = {
  CLeft,   GLeft,
  CRight,  GRight,
  CUp,     GUp,
  CDown,   GDown,
  CB,      GJump,
};
```

Figure 2: Controller and game predicates, and a mapping from one to the other.

statements. Using enumerations, constant variables, or preprocessor macros a set of constants like $\mathbb{G} = \{left, right, up, down, jump\}$ shall be represented as *unsigned integers*—series of bits—such that each element in $\mathbb{G}$ is assigned a bit.

Then, a single machine word can store any subset of the $\mathbb{G}$ provided that $\mathbb{G}$ has at most as many elements as there are bits in a machine word. This might seem like a severe restriction but so long as we are only concerned with digital inputs—and for analog inputs it will often suffice to treat an axis as one or two digital inputs—we will often find that the number of inputs on a controller does not exceed the number of bits in a machine word of the architecture. Early controllers—for the Nintendo Famicom/NES, NEC PC-Engine/TurboGrafx-16, Sega SG-1000/Master System series—had at most[3] eight digital inputs in the form of a four-way directional pad, two main face buttons and two menu buttons, which just so happen to fit into the 8-bit registers featured in their respective console's CPUs.

As the number of inputs on a controller grows, so does the number of bits in available registers. With few exceptions, all popular video game consoles satisfy this criterion: If we simplify—and map an analog trigger or face button to one digital input, and an axis on an analog stick to two—the controllers of the Nintendo 64, Gamecube, Wii and Wii U, Switch, Sega's Saturn and Dreamcast, the Sony PlayStation, and Microsoft Xbox series, as well as various handhelds, usually do not exceed a maximum of 25 inputs—well below the usual register size of 32 bits. Many other, more exotic, offerings follow this trend.

On the rare occasion that this is not the case—for example, while the Super Famicom/Super Nintendo Entertainment System's Ricoh 5A22 CPU has an 8-bit accumulator register, there are 14 inputs on the standard controller—it may be necessary to resort to distributing values to multiple data words and additional bit shifting operations. In certain cases we might even be able to circumnavigate this problem, e.g., when we content ourselves with only applying the method to some inputs—as in, having in-game actions fixed for a game pads directional pad, allowing only for face buttons or triggers to be mapped freely—so as to bring the number of inputs down to a number which may fit inside a single register.

Assuming a situation in which all inputs are indeed unambigously mapped to their corresponding bits inside a single CPU register—and similarly so for all in-game actions—a predicate is built from multiple sets, each of which is constructed using *bitwise or* operations for all elements which make up a given set, and a controller-game mapping may be represented as an array of machine words, with each two-word pair encoding that we wish to map one constant to another. Mapping a set of sets is then just a matter of iterating through the source sets: For each, we start out with an "empty" destination set represented by a variable where all bits are zero, and then, for each bit that is set in the source we use *bitwise and* operations to add any corresponding constants to the result. Fig. 2 shows how one could encode predicates and mappings in C; the simple way the data are laid out in memory should make for easy workability, even in an assembly language.

Fig. 3 shows how predicates are mapped from one set of constants to another, and how to test whether one predicate is a *sub-predicate* of another. Note that all constants are mere integer constants and the values chosen for distinct sets of constants may overlap. These two functions, and a way to construct a mapping, are all that is needed to let a user choose a button mapping and test whether it will provide a satisfactory configuration. The method can be expanded to take into account multiple predicates which act in concert, or to determine which, if any, problems there are with a given mapping.

The predicates themselves are either created at build time—by hand or automated—or, in the case of the controller used being only known at run time, through interaction with the user. We may ask the user to try to press certain button combinations to determine whether they are suitable, or resort to ready-made building blocks which cover a wide range of controller types, as we shall discuss in section 7.

---

[3]For the NEC PC-Engine/TurboGrafx-16 there were both a more commonly used controller with two face buttons—eight digital inputs—and a later six-button model—12 inputs. We ignore the six-button version here, but handling a few additional buttons should only be a minor inconvenience.

```
void map(wt src[], wt src_n, wt src_m,
  wt mapp[], wt mapp_n, wt dest[]) {
  for (wt i = 0; i < src_n; i++) {
    dest[i] = 0;
    for (wt b = 0; b < src_m; b++) {
      if (src[i] & (1<<b)) {
        for (wt j = 0; j < mapp_n; j++) {
          if (mapp[j * 2] & (1<<b)) {
            dest[i] |= mapp[j * 2 + 1];
          }}}}}}

bool subset(wt sub[], wt sub_n,
  wt super[], wt super_n) {
  for (wt i = 0; i < sub_n; i++) {
    bool found = false;
    for (wt j = 0; j < super_n; j++) {
      if (sub[i] == super[j]) {
        found = true;
        break;}}
    if (!found) {
      return false;}}
  return true;}
```

Figure 3: C code for mapping predicates and checking for subsets.

## 6 Propositional logic

In addition to the interpretation as a predicate, or a set of sets of constants for which the predicate is true, we can also view what we know about—or demand of—controllers as logical formulas.

If for some set $\mathbb{R}$ we know that it is a subset of $\mathbb{C}$—i.e., that all elements of $\mathbb{R}$ must also be in $\mathbb{C}$—then it follows naturally that the set $\mathbb{R}$ can also be represented as a *propositional formula* of the following form:

$$\bigwedge_{c \in \mathbb{C}} \begin{cases} c & \text{if } c \in \mathbb{R} \\ \neg c & \text{otherwise} \end{cases}$$

This formula contains a literal for each element in $\mathbb{C}$—negated for those which are not in $\mathbb{R}$. For example, if $\mathbb{C}$ were $\{a, b, c\}$ and $P(\{a, c\}) = true$ and $P$ gave $false$ for all other sets, the corresponding formula would be $a \wedge \neg b \wedge c$.

Most of the predicates we are concerned with map multiple sets to $true$. Let us define those sets which some predicate $P$ maps to $true$ as sets $\mathbb{X}_1, \mathbb{X}_2, \ldots, \mathbb{X}_n$. Then $P(\mathbb{R}) = true$, if and only if $\mathbb{R}$ is equal to $\mathbb{X}_i$ for some $i$, i.e., $\bigvee_{i=1}^{n} \mathbb{R} = \mathbb{X}_i$.

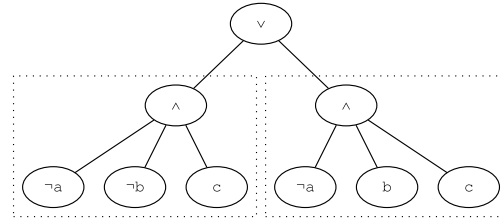

Figure 4: Parse tree of our running example propositional formula.

By combining these two arguments we conclude that for predicates on sets, as we have described them, there is always a corresponding propositional formula.

**Definition 7.** Let $\mathbb{C}$ be a set of scalar values, $P : \mathcal{P}(\mathbb{C}) \mapsto \mathbb{B}$ be a predicate, and $\mathbb{X}_i$ for $0 < i \le n$ for some $n \ge 1$ be the sets for which $P(\mathbb{X}_i) = true$.

Then we define the *propositional formula* corresponding to $P$ as

$$\bigvee_{i=1}^{n} \left( \bigwedge_{c \in \mathbb{C}} \begin{cases} c & \text{if } c \in \mathbb{X}_i \\ \neg c & \text{otherwise} \end{cases} \right)$$

Each $\mathbb{X}_i$ corresponds to an assignment of $true$ and $false$ to propositional variables that makes this formula $true$. There are two ways to read this formula, which together show the connection between predicate and propositional formula: The formula is in *disjunctive normal form*, that is to say, it is a disjunction of conjunctions.

Let us consider as a running example the formula $(\neg a \wedge \neg b \wedge c) \vee (\neg a \wedge b \wedge c)$. Fig. 4 shows this formula as a tree of $\wedge$, $\vee$ and *literal* nodes.

Reading the formula, we see that it contains two clauses—in Fig. 4 their corresponding subtrees are contained in dotted frames—which are connected by a $\vee$ node—the root of the tree. The formula as a whole evaluates to $true$ if one or the other clause is $true$, or if both are. Under the interpretation as a predicate, this says that $P(\mathbb{X}) = true$ if and only if $\mathbb{X}$ is the set described by either one clause *or* the other—by either the left *or* the right framed subtree. What unites these two interpretations is that they both talk about *options* which are to be iterated over, and, once a positive result is found, the result of the whole query is positive. If all options are exhausted without finding the correct set—or a clause that is $true$ under some assignment—the result is negative.

It remains to interpret the clauses—which are conjunctions of literals, and represent sets. As a propositional formula they tell us not only what has to be $true$ but also what has to be $false$. In logic formulas this needs to be spelled out explicitly because, for example, omitting $\neg a$ from the first—or left-hand—clause would change its meaning from "$a$ and $b$ shall be $false$ and $c$ shall be $true$" to "$b$ shall be $false$ and $c$ shall be $true$, but $a$ may be either." Here the common trait is that a set and a conjunction of literals both exhaustively describe a structure by describing its parts; one describes which variables are assigned $true$, and which are assigned $false$; the other describes which constants are *in* a set, and which are *not*.

With sets, checking whether the requirements of a game are covered by the facilities of a controller amounts to checking whether the predicate set of the game is a subset of the *mapped* predicate set of the controller. Algorithmically this is achieved by checking for each set of the game's predicate whether it is also one of the controller's sets.

The corresponding method for propositional formulas works similarly: A controller's facilities are converted to a single disjunctive normal form, while a game's requirements take the form of a *set of conjunctions*. For each such conjunction we need to check whether its logical conjunction with the controller's formula is *satisfiable*, that is to say, whether there is a truth assignment which satisfies—evaluates to $true$—both at the same time. Finding such an assignment for each conjunction representing a game requirement corresponds to confirming that a game's predicate set is a subset of that of a controller.

If we are now sufficiently convinced that these two representations are indeed interchangeable, we might still ask why we should go to the trouble of converting from one to the other. Our answer is that propositional formulas conveniently lend themselves to be minimized—to various degrees and with various amounts of effort—via algorithms such as those described by Quine and McCluskey [7, 6, 5]. The method described in the previous paragraph gives the same result for all versions of a controller's formula, but a minimized formula has some potential for reducing computation time.

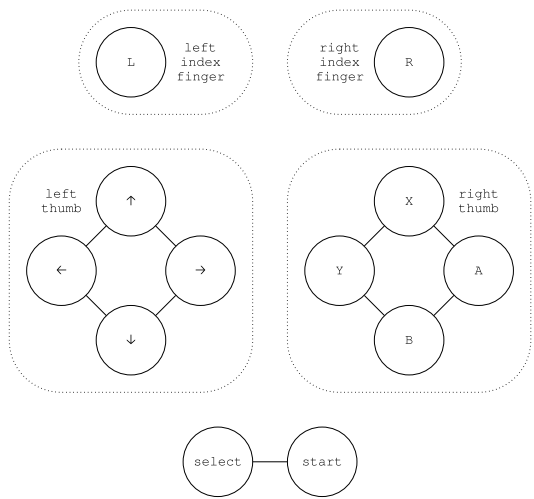

Figure 5: One way to model the SFC/SNES controller.

## 7 Creating predicates

The structure of our predicates may not directly reflect the way we naturally think about game controls. In this section, let us try to make the process of creating them more intuitive.

Consider Fig. 5. It depicts all inputs of Nintendo's SFC/SNES controller in roughly the same layout as on the actual device. Edges are meant to represent that two inputs may be pushed by a single finger simultaneously. For each connected component of the graph the predicate set is just the set of edges between the component's nodes. Components which are activated using a single human finger are shown in dotted frames. The $select$ and $start$ buttons can be pushed using the right or left thumb, and during gameplay they usually do not need to pushed at the same time as any other buttons, so they are not discussed further. These predicates—the ones corresponding to framed subgraphs—can then be combined using the $simultaneously$ combinator introduced in section 4.

In the situation that the physical controller is unknown at the time of development we want to offer a simple procedure to the user. Two approaches

come to mind:

A large proportion of controllers seem to be made from just a few recurring "building blocks." Directional pads and analog sticks are probably the most common examples. "Triggers" or "shoulder buttons" usually come in symmetric pairs, or symmetric pairs of two.[4]

As for "face buttons" the most common layouts are shown in Fig. 6. As before, edges signify that two buttons can easily be pushed simultaneously. Two-button layouts like the one shown are used not only on Nintendo's Famicom/NES, but on many so-called "8-bit" generation systems, such as the Sega Master System, NEC PC-Engine and MSX home computers. On contemporary system, a four-button layout is most common. It is in use on Nintendo's Super Famicom/SNES controller and on the more recent Switch Pro Controller. Sony and Microsoft use similar layouts with different button specifiers. Controllers for use on personal computers often follow the design of console offerings—meaning again, four buttons. Six-button configurations are used by a number of Sega systems, and some are offered as alternatives to the standard Xbox and PlayStation controllers. We can let a user choose among these layouts and then combine ready-made predicates accordingly.

Button combinations including menu buttons are rather uncommon, but one will—again—often encounter the same configurations and these can then be handled in the way shown for face buttons.

Alternatively, if the number of required button combinations is small, we may also ask the user to specify which buttons are easy to be activated—e.g., by pressing button pairs, until enough pairings have been discovered—and then build predicates from that information.

## 8 Conclusion & Future Work

The method shown is a compromise between a simple design and expressive power. It is leaning towards simpleness; not every desired quality may be expressed as simple sets of sets. Some enhancement are straight-forward extensions, such as allowing ordered lists of sets and scalars in a predicates domain. Using sets we can only express that, for example, buttons shall have the *same* size, or that a set of buttons shall have some quality *in common*; if we want to express that *one* button shall be larger, or smaller than *another*, or to the left of it, we need these slightly more complex structures.

From a somewhat haphazard survey of games which fall, for the most part, into the "16 and 32 bit" generation of console video games we draw some conclusions:

There is evidence that even when a system's standard controller only had, say, three buttons, some consideration went into which actions to map to which buttons and also into what options to offer to the player. Some common approaches follow.

All possible mappings: At least for face buttons, when there is only a limited quantity, a number of games offer all possible combinations. For the Sega Mega Drive's three-button controller there are only six ways to map three in-game actions to buttons. Games which need fewer buttons than are available sometimes map multiple buttons to the same in-game action, which could be said to represent multiple layouts in one.

A range of control schemes: Other times the set of control schemes is limited to just a few options. In The Revenge of Shinobi on Mega Drive/Genesis[5] all control schemes follow the rule that the special *Ninjutsu* attack is never mapped to the $B$ button which is located in-between $A$ and $C$. This results in the $jump$ and $attack$ buttons always being mapped to adjacent buttons and can be expressed using predicates. These and other examples might suggest that requirements, such as that two buttons should be adjacent, played a role in deciding which control schemes to allow and which not to offer at all. In other cases it is possible to discern some basic pattern that could also be expressed using the predicate method. Examples include special attack buttons, which are rarely used—as the number of times a special attack may be per-formed is limited—and do not need to be pushed in combination with any other buttons—and thus may be set to a fixed button—while what can be considered main actions—jumping, attacking, running—are offered in a few variations.

---

[4]Exceptions are Nintendo's Nintendo 64 and GameCube controllers; the Nintendo 64's controller is designed to be held in three different configurations: its Z-trigger takes the place of the left trigger in most games, some games use the L-trigger *instead*, and on rare occasions L and Z are used; on the GameCube's controller the Z-trigger is rather small and serves as a secondary right trigger.

[5]Known in Japan as The Super Shinobi. The same control schemes are found in Shinobi III: Return of the Ninja Master, or The Super Shinobi II in Japan.

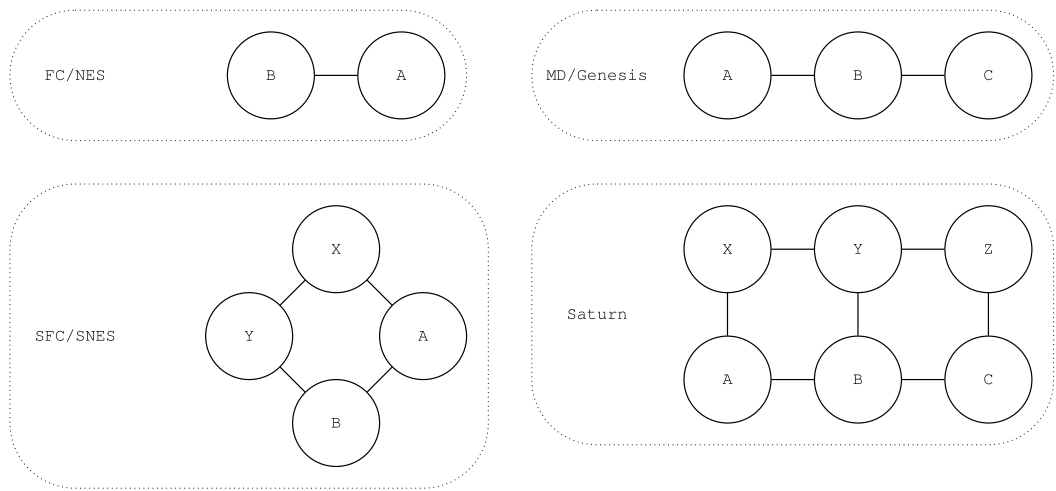

Figure 6: Common "face button" layouts, simplified from their appearances on Nintendo Famicom/NES, Super Famicom/SNES, Sega Mega Drive/Genesis and Saturn controllers.

Mapping individual buttons/in-game actions: Some games allow all—or sometimes only a subset of—buttons to be mapped to a single function. This may lead to configurations that do not allow gameplay as intended, such as mapping all functions to a single button in Gradius III or not mapping any buttons to in-game actions in Mega Man X. In contrast, some games do not allow a user to exit the options menu without configuring a button for each necessary in-game action. We also encountered a select few games which allow multiple functions to be mapped to the same button.

Most often, the control schemes mapped individual buttons to individual in-game actions. Some games which need fewer buttons than are available allow multiple buttons for the same function. Infrequently, control schemes feature a button that triggers a button combination that could also be executed by pressing multiple buttons, which probably is due to the fact that arcade games ported to home consoles are often designed for far fewer buttons than are available on a home console controller.[6]

This brings us to another case which was influential during our research: Treasure's Radiant Silvergun in the arcade version is played using an eight-way digital joystick and three buttons. Each button is assigned a different type of shot, and on a standard arcade cabinet each of these buttons can be activated using a different finger of the right hand. That is to say, one may push any combination of these three buttons at any time. This is in fact *required* to use all functions the game offers: Not only can *any two* of the three attacks be combined to form another set of three attacks, pushing all three buttons at once is used to activate a short-range sword type weapon, or—if enough energy has been collected—unleash a special attack. In effect, seven different in-game actions are controlled by just three buttons. While arcade sticks are also available for home use, the Sega Saturn version features the following adaptation for regular handheld game controllers: In addition to $A$, $B$ and $C$ buttons—mapped to the game's three primary shot types, as in the arcades—$X$, $Y$ and $Z$ map to $B \wedge C$, $A \wedge C$ and $A \wedge B$, respectively. This allows the player, first, to use the $X$, $Y$ and $Z$ buttons to activate the three combined attack options, and second, the buttons in the upper row—i.e., $X$, $Y$, $Z$—are arranged so that, together with the button below them—$A$, $B$ and $C$—they always give the full set of inputs, namely $A \wedge (B \wedge C) = B \wedge (A \wedge C) = C \wedge (A \wedge B)$. In addition, the Saturn controller's right trigger, $R$, is mapped to the full set $A \wedge B \wedge C$, that is, the seventh weapon. Would this not be so, it would be at best be uncomfortable—and at worst, impossible for some players—to push the combination of the $A$ and $C$—but not $B$—buttons using just the right thumb. For reference see Fig. 6 of the Saturn's six-button layout. A later port to Microsoft's Xbox Live Arcade ecosystem features similar button mappings.

---

[6]The standard JAMMA connector—by the Japan Amusement Machine and Marketing Association—defines just three action buttons; extensions exist, which feature more buttons.

There are a few more directions in which to continue exploring the presented method and related questions:

A more refined hierarchy of controller "building blocks" and ways to combine them seem to be an area worthwhile of further research: Most controllers fall into one of a few categories, but there are multiple answers to the question of whether—and where—to divide a controller into subordinate parts.

Rather than seeking out those button combinations which are easy to press or hold, we could also pursue the converse approach: seeking out hard to press button combinations on purpose. With careful design, we could imagine situations such as a climbing mini-game where hard to hold button combinations simulate hard to reach climbing holds.

Additionally, we wonder, can this approach be utilized to *adapt* given control schemes—for the injured, or the disabled? If a button, or button combination, is out of reach for some users, is there a way we can automatically generate and suggest alternatives?

Further insight may also be gained from more thorough and extensive surveys of what control schemes and ways of configuring controllers are used in practice: What is the degree of freedom offered? What nonsensical configurations are allowed and disallowed? In what ways is the user aided in finding a satisfactory button mapping? In case there are more buttons available than are needed, is it a matter of personal taste which buttons one uses or are there universal categorizations into different tiers that buttons fall into?